-----------------------------------------------------------------------------

\documentstyle[12pt]{article}
\begin{document}
\newcommand{\frad}{\displaystyle\frac}
\newcommand{\ba}{\begin{array}}
\newcommand{\ea}{\end{array}}

\rightline{BIHEP-TH-94-32}

\vspace{1.5cm}
\centerline{\bf\Large Reliability of the Estimation of CP Asymmetries}
\centerline{\bf\Large for Nonleptonic $B^{0} - \bar{B}^{0}$ Decays}
\centerline{\bf\Large into Non-CP-Eigen states}

\vspace{2.5cm}

\begin{center}
{\bf Dongsheng Du$^{a,b}$, Xiulin Li$^{c}$, and Zhenjun Xiao$^{a,d}$}
\end{center}
a, China Center for Advanced Science and Technology (World Lab.)

P.O. Box 8730, Beijing 100080, P.R.China\\
b, Institute of High Energy Physics, Chinese Academy of Sciences,

Beijing 100039, P.R.China * \\
c, Department of Physics, Hangzhou Teacher's College,

Hangzhou 310012, P.R.China * \\
d, Department of Physics, Henan Normal University,

Xinxiang 453002, P.R.China *

\begin{abstract}
CP asymmetries for two-body-nonleptonic $B^{0} - \bar{B}^{0}$ decay into
non-CP-eigen states are calculated using two different methods: (i)
Bauer, Stech, and Wirbel factorization method to compute the decay amplitudes
directly; (ii) using $B^{0}, \bar{B}^{0}$ decay amplitude ratios to avoid the
direct computation of the decay amplitudes. The comparison of the results are
made. The conclusion is presented.
\end{abstract}

\vspace{0.5cm}

PACS number(s): 11.30.Er, 13.25.tm

*Permanent Address
\newpage
\section*{I. Introduction}

{\hskip 0.6cm}The CP asymmetries for $B^{0} - \bar{B}^{0}$ two-body nonleptonic
decays have been systematically estimated$^{[1]}$. As proved in Ref.[1]
(see its Appendix), for CP-eigen states f the amplitude ratios
$$
\zeta = 
{A(\bar{B}^{0} \rightarrow f)}/{A(B^{0} \rightarrow f)} ~,
{}~~ \bar{\zeta} = 
{A(B^{0} \rightarrow \bar{f})}/{A(\bar{B}^{0} \rightarrow
\bar{f})}
$$
depend only on KM matrix elements. But if the final state f is not a
CP-eigen state, $\zeta, ~\bar{\zeta}$ are not pure KM factors. In the
estimation of CP asymmetries in Ref.[1], the pure KM factor approximation
for $\zeta$ and $\bar{\zeta}$ is also used for non-CP-eigen states, such as
$D^{\pm}\pi^{\mp}$ etc. . But how reliable these estimations are? In this
short article, we first compute the CP asymmetries for $B^{0} - \bar{B}^{0}$
decays into non-CP-eigen states using
Bauer, Stech, Wirbel factorization
method$^{[2]}$. Then we compare these results with those by using the amplitude
ratios $\zeta$ and $\bar{\zeta}$. In section II we present all the results of
the two different methods. Section III devotes to the discussions and
conclusions.

\section*{II. Computation of the partial-decay-rate asymmetries}

{\hskip 0.6cm}For simplicity of comparison, we consider only incoherent
$B^{0}_{d} - \bar{B}^{0}_{d}$ mesons. For instance, sitting on the $Z^{0}$
resonance, $b\bar{b}$ pairs will be produced in the form of $B^{0}_{d}
B^{-}_{u}$ and $\bar{B}^{0}_{d} B^{+}_{u}$. Here $B^{0}_{d}, ~\bar{B}^{0}_{d}$
are produced incoherently and observing the charge of $B^{-}_{u}(B^{+}_{u})$
would confirm the decayed neutral meson to be $B^{0}_{d}(\bar{B}^{0}_{d})$.
In the decays of incoherent $B^{0}_{d} - \bar{B}^{0}_{d}$ mesons, we can
define the CP asymmetry parameter as
$$
a_{f}(t) = \frac{\Gamma(B^{0}_{d, Phys.}(t) \rightarrow f) - \Gamma(B^{0}_{d,
Phys.}(t) \rightarrow \bar{f})}{\Gamma(B^{0}_{d, Phys.}(t) \rightarrow f)
+ \Gamma(B^{0}_{d, Phys.}(t) \rightarrow \bar{f})} \eqno (2.1)
$$
where
$$
\ba {rl}
|B_{d,Phy.}^0(t)> & =f_+(t)|B_d^0>+\frad{q}{p}f_-(t)|\bar{B}_d^0>\\
|\bar{B}_{d,Phy.}^0(t)> & =\frad{p}{q}f_-(t)|B_d^0>+f_+(t)|\bar{B}_d^0>
\ea \eqno (2.2)
$$
$$
\ba {rl}
|B_L> & =p|B_d^0>+q|\bar{B}^0_d>\\
|B_H> & =p|B_d^0>-q|\bar{B}^0_d>\\
\ea \eqno (2.3)
$$
$$
\ba {rl}
f_{\pm}(t) & =\frad{1}{2}(e^{-i\lambda_Lt} \pm e^{-i\lambda_Ht})\\
\lambda_{L,H} & =m_{L,H}-\frad{i}{2}\Gamma_{L,H}
\ea \eqno (2.4)
$$
Integrating with time t from zero to infinity we can get the integrated CP
asymmetry
$$
{\cal A}_{f} = \int^{\infty}_{0} dt ~~ a_{f}(t) \eqno (2.5)
$$
Now we first compute the asymmetry parameter ${\cal A}_{f}$ in
Bauer, Stech, Wirbel (BSW) scheme.

\vspace{0.5cm}

We take $B^{0}_{d} \rightarrow D^{+}\pi^{-}$ as an example for the purpose of
illustration.

\vspace{0.5cm}

For $B^{0}_{d} \rightarrow D^{-} \pi^{+}$, the effective Hamiltonian is$^{[2]}$
$$\ba {rl}
{\cal H}_{eff} & =\frad{G_F}{\sqrt{2}}V_{cb}^*V_{ud}
\{a_1[\bar{b}\gamma_{\mu}(1-\gamma_5)c]_H
[\bar{u}\gamma^{\mu}(1-\gamma_5)d]_H\\
&~~~ +a_2[\bar{b}\gamma_{\mu}(1-\gamma_5)d]_H
[\bar{u}\gamma^{\mu}(1-\gamma_5)c]_H\}+h.c.
\ea \eqno (2.6)
$$
Neglecting the contribution of the exchange diagram we have
$$
<D^-\pi^+|{\cal H}_{eff}(0)|B^-_d>
\cong \frad{G_F}{\sqrt{2}}if_{\pi}a_1p_{\pi}^{\mu}
<D^-|\bar{b}\gamma_{\mu}(1-\gamma_5)c|B_d^0> \eqno (2.7)
$$
where
$$
<\pi^+|\bar{u}\gamma^{\mu}(1-\gamma_5)d|0>=if_{\pi}p_{\pi}^{\mu}
$$
has been used.

Using
$$\ba {rl}
<D^-|\bar{b}\gamma_{\mu}(1-\gamma_5)c|B_d^0> & =
[(p_B+p_D)-\frad{m_B^2-m_D^2}{q^2}q]_{\mu}F_1(q^2)\\
& +\frad{m_B^2-m_D^2}{q^2}q_{\mu}F_0(q^2)
\ea \eqno (2.8) $$
and $q_{\mu} = (p_{B} - p_{D})_{\mu}$, we finally get
$$
<D^-\pi^+|{\cal H}_{eff}(0)|B^-_d>\approx
\frad{G_F}{\sqrt{2}}V_{cb}^*V_{ud}if_{\pi}a_1(m_B^2-m_D^2)F_0^{BD}(m_{\pi}^2)
\eqno (2.9)$$
Similarly we can get
$$
<D^-\pi^+|{\cal H}_{eff}(0)|\bar{B}^-_d>\approx
\frad{G_F}{\sqrt{2}}V_{ub}V_{cd}^*if_Da_1(m_B^2-m_{\pi}^2)F_0^{B\pi}(m_D^2)
\eqno (2.10) $$
Also
$$
\ba {rl}
<D^+\pi^-|{\cal H}_{eff}(0)|B^-_d>\approx &
\frad{G_F}{\sqrt{2}}V_{ub}^*V_{cd}if_{D}a_1(m_B^2-m_{\pi}^2)F_0^{B\pi}(m_D^2)\\
<D^+\pi^-|{\cal H}_{eff}(0)|\bar{B}^-_d>\approx &
\frad{G_F}{\sqrt{2}}V_{ud}^*V_{cb}if_{\pi}a_1(m_B^2-m_D^2)F_0^{BD}(m_{\pi}^2)
\ea \eqno (2.11)$$
The CP asymmetry parameter
$$
a_{D^-\pi^+}(t)=\frad
{|<D^-\pi^+|{\cal H}_{eff}|B^0_{d,Phys.}(t)>|^2
-|<D^+\pi^-|{\cal H}_{eff}|\bar{B}^0_{d,Phys.}(t)>|^2}
{|<D^-\pi^+|{\cal H}_{eff}|B^0_{d,Phys.}(t)>|^2
+|<D^+\pi^-|{\cal H}_{eff}|\bar{B}^0_{d,Phys.}(t)>|^2}
$$
After the time integration we have
$$
{\cal A}_{D^-\pi^+}=-\frad
{0.514f_\pi f_Dm_B^2(m_B^2-m_D^2)x_dIm(\frad{
V_{tb}^*V_{td}V_{cb}V_{ub}V_{ud}^*V_{cd}^*}{V_{tb}V_{td}^*})}
{0.476f_{\pi}^2(m_B^2-m_D^2)^2|V_{ud}^*V_{cb}|^2(2+x_d^2)
+0.138f_D^2m_B^4x_d^2|V_{cd}^*V_{ub}|^2}
\eqno (2.12)
$$
where the values of the form factors are take from Ref.[2].

\vspace{0.5cm}

For the KM factors, we use the Wolfenstein parametrization$^{[3]}$
$$
V=\left(
\ba {lcr}
V_{ud}&V_{us}&V_{ub}\\
V_{cd}&V_{cs}&V_{cb}\\
V_{td}&V_{ts}&V_{tb}
\ea \right)
=\left(\ba {lcr}
1-\frad{1}{2}\lambda^2&\lambda&A\lambda^3(\rho-i\eta)\\
-\lambda&1-\frad{1}{2}\lambda^2&A\lambda^2\\
A\lambda^3(1-\rho-i\eta)&-A\lambda^2&1
\ea \right)
\eqno (2.13) $$
{}From the updated fit [4]
$$
A\sim 0.84, \lambda\sim 0.22, \sqrt{\rho^2+\eta^2}\sim 0.36 \eqno (2.14)
$$
The values of $\rho$ and $\eta$ depend on the top quark mass $m_{t}$ and
$f_{B_{d}} \sqrt{B_{B_{d}}}$. For the purpose of illustration, we take
$m_{t} \sim 174 GeV$, $f_{B_{d}} \sqrt{B_{B_{d}}} \sim 180 GeV$ and [4]
$$
\rho = -0.05, ~~~~ \eta = 0.33 \eqno (2.15)
$$
The other parameters are taken as
$$
\begin{array}{rl}
f_{\pi} & = 0.13 GeV, ~~~ f_{K} =f_{D} = 0.16 GeV \\
m_{B_{d}}& = 5.28 GeV, ~~m_{D^{+}} = 1087 GeV,\\
x_{d} & \sim 0.7  \end{array} \eqno (2.16)
$$
Substituting all these parameters into Eq.(2.12), we get the time-integrated
CP asymmetry
$$
{\cal A}_{D^{-}\pi^{+}} = - 5.0 \times 10^{-3} \eqno (2.17)
$$

For the final states involved vector mesons, we use
$$
< 0 \mid V_{\mu} \mid V > = \lambda_{V} m^{2}_{V} \epsilon_{\mu}(V) \eqno
(2.18)
$$
and take
$$
\lambda_{D^{*}} = 0.14, ~~ \lambda_{\rho} = 0.24 \eqno (2.19)
$$
Thus, we can use BSW method to compute the asymmetries for different processes.
We
put all these results in Table I. Note that for the final state f for which
$B^{0}_{d} \rightarrow f$ can occur but $\bar{B}^{0}_{d} \rightarrow f$
cannot, there will be no CP asymmetry.

\vspace{0.5cm}

In order to compare these results with those by use of amplitude ratios, we
compute
the same CP asymmetries by use of the amplitude ratios like in Ref.[1] but use
the
same KM parameters as in BSW method. The results are presented also in Table I
(denoted by AR method).

\section*{III. Discussions and conclusions}

{\hskip 0.6cm}In Table I, we show both the results of BSW and AR (Amplitude
Rations) methods. From that table we can see that for most of the processes
listed
there, the asymmetries by both methods are very close to each other. For very
few processes, such as $f = D^{-}\rho^{+}$, there are large discrepancy but at
most
a factor 2 difference. So on the whole they agree to each other. At least the
order of magnitudes of the CP asymmetries is reliable. But, if we want to use
these CP asymmetries to make a precision test of the standard model, it is not
good
at all. For other purposes, the CP asymmetries computed by both BSW and AR
still can be used. We must remind the reader that we are now talking about
the non-CP-eigen states. If the final state is a CP-eigen state, the AR
method can give a reliable prediction of the CP asymmetry.

\vspace{0.5cm}

This work is supported in part by the National Natural Science Foundation of
China and the State Commission of Science and Technology of China.

\section*{Table Captions}

\noindent
Table I. Asymmtretries in BSW and AR scheme. Here BSW means Bauer, Stech,
Wirbel method [2], AR means Amplitude Ratio method [1].

\newpage
{\hskip 2.5cm}{Table I}

\begin{center}
\begin{tabular}{r|cc} \hline
Process & ${\cal A}_f(BSW)$ & ${\cal A}_f(AR)$\\  \hline
$B_d^0\rightarrow D^-\pi^+$ &$-5.0\times 10^{-3}$ &$-6.6\times 10^{-3}$\\
$D^+\pi^-$ &$-2.6\times 10^{-2}$ &$-3.4 \times 10^{-2}$ \\
$D^{*-}\pi^+$ &$9.1 \times 10^{-3}$ &$6.8 \times 10^{-3}$ \\
$D^{-}\rho^+$ &$2.3 \times 10^{-3}$ &$6.8 \times 10^{-3}$ \\
$\pi^{-}\rho^+$ &0.24 &0.37 \\
$D^{*-}D^+$ &-0.34 &-0.25 \\
$D^{-}D^{*+}$ &-0.15 &-0.26 \\
$\pi^{-}D^{*+}$ &$4.6 \times 10^{-2}$ &$3.4 \times 10^{-2}$ \\
$\pi^+\rho^-$ &0.48 &0.37 \\
$\bar{D}^0\pi^0$ &$-7.0 \times 10^{-3}$ &$-6.8 \times 10^{-3}$ \\
$\bar{D}^0\eta$ &$-7.0 \times 10^{-3}$ &$-6.8 \times 10^{-3}$ \\
$\bar{D}^0\eta^,$ &$-7.0 \times 10^{-3}$ &$-6.8 \times 10^{-3}$ \\
$\bar{D}^{*0}\pi^0$ &$-6.9 \times 10^{-3}$ &$-6.8 \times 10^{-3}$ \\
$\bar{D}^{*0}\eta$ &$-6.9 \times 10^{-3}$ &$-6.8 \times 10^{-3}$ \\
$\bar{D}^{*0}\eta^,$ &$-6.9 \times 10^{-3}$ &$-6.8 \times 10^{-3}$ \\
$\bar{D}^{0}\rho^0$ &$-7.1 \times 10^{-3}$ &$-6.8 \times 10^{-3}$ \\
$\bar{D}^{0}\omega^{0}$ &$-7.1 \times 10^{-3}$ &$-6.8 \times 10^{-3}$ \\
$D^0\pi^0$ &$-3.5 \times 10^{-2}$ &$-3.4 \times 10^{-2}$ \\
$D^0\eta$ &$-3.5 \times 10^{-2}$ &$-3.4 \times 10^{-2}$ \\
$D^0\eta^,$ &$-3.5 \times 10^{-2}$ &$-3.4 \times 10^{-2}$ \\
$D^{*0}\pi^0$ &$-3.5 \times 10^{-2}$ &$-3.4 \times 10^{-2}$ \\
$D^{*0}\eta$ &$-3.5 \times 10^{-2}$ &$-3.4 \times 10^{-2}$ \\
$D^{*0}\eta^,$ &$-3.5 \times 10^{-2}$ &$-3.4 \times 10^{-2}$ \\
$D^0\rho^0$ &$-3.6 \times 10^{-2}$ &$-3.4 \times 10^{-2}$ \\
$D^0\omega^0$ &$-3.6 \times 10^{-2}$ &$-3.4 \times 10^{-2}$ \\
\hline
\end{tabular}
\end{center}
\end{document}